\documentclass[12pt]{article}

\usepackage[utf8]{inputenc}
\usepackage[T1]{fontenc}
\usepackage[margin=1in]{geometry}
\usepackage{graphicx}
\usepackage{amsmath}
\usepackage{amssymb}
\usepackage{booktabs}
\usepackage{longtable}
\usepackage{authblk}
\usepackage{caption}
\PassOptionsToPackage{hyphens}{url}
\usepackage[colorlinks=true, linkcolor=blue, citecolor=blue, urlcolor=blue]{hyperref}
\usepackage{xurl}
\usepackage{array}
\usepackage{hanging}

\title{reGRAF: a global MPAS reforecast data set with \\
       convection-allowing refinements over \\ the US and Europe}

\author[1]{Thomas M. Hamill}
\author[2]{Akshay Subramaniam}
\author[1]{Montgomery Flora}
\author[2]{Raghu Raj Prasanna Kumar}
\author[2]{Karthik Kashinath}
\author[2]{Carl Ponder}
\author[2]{Tao Ge}
\author[2]{Sepideh Khajehie}
\author[1]{John Wong}
\author[1]{Brett Wilt}
\author[1]{Peter Neilley}

\affil[1]{The Weather Company, Atlanta, Georgia, USA}
\affil[2]{NVIDIA, Santa Clara, California, USA}

\date{\today}

\begin{document}
\maketitle
\raggedright
\setlength{\parskip}{6pt}

\begin{abstract}
NVIDIA and The Weather Company (TWC) have generated a data set of
reforecasts from TWC's GRAF (Global high-Resolution Atmospheric
Forecasting) model, a version of the National Center for Atmospheric
Research (NCAR) Model for Predictions Across Scales
(\href{https://ncar.ucar.edu/what-we-offer/models/model-prediction-across-scales-mpas}{\underline{MPAS}}).
GRAF is global, but the configuration for this reforecast had a mesh
refinement to \textasciitilde4 km over the US, Caribbean Basin, and
Europe, and 15 km elsewhere. This model was designed to run much of the
computation on graphical processing units (GPUs), with this development
assisted by NVIDIA.

The 1836 reforecast cases (\textasciitilde5 years) were generated from
ECMWF reanalyses (ERA5) for selected initial condition dates spanning
more than 20 years, 2004--2024. These dates of the chosen initial
conditions were mostly selected based on high-impact weather in the
contiguous US (CONUS) and Caribbean. Sampling in this way, the
reforecast spanned a wider range of interesting, high-impact weather
scenarios than had we performed five contiguous years of once-daily
reforecasts. The reforecast still provides many samples in
non-precipitating regions with more ordinary weather.

GRAF reforecasts were mostly run to $+$27 h lead time, assuming a 3-h spin-up
followed by a full diurnal cycle. Data were saved in zarr format
on the native model vertical coordinate. Most fields were archived at
15-min intervals, though several precipitation variables were saved at
5-min cadence. Data are made publicly available to all through Amazon
Web Services' Open-Data Initiative at https://registry.opendata.aws/graf-reforecast.
\end{abstract}

\section{Introduction}

Many organizations in the weather enterprise seek to provide
ever-improving predictions, from which customers can make better and
better decisions. Numerical weather prediction (NWP) is the underpinning
technology behind accurate predictions of weather changes. These
improvements have accumulated at a rate of about one day a decade, i.e.,
a four-day forecast now is as accurate as a three-day forecast produced
a decade ago. The slow accumulation of skill, which translates into
improved products and services, represents a ``quiet revolution'' in
weather prediction (Bauer et al. 2015).

Within the last few years and with the advance of artificial
intelligence, a radically different approach to NWP has been developed.
The new models are \emph{data driven}; they do not represent a complex
human codification of the physical laws of motion, parameterized
processes, and the interactions between state components. Instead,
comparatively simple neural-network models are trained. In a common
method of coding these models for global weather prediction, this provides a
sophisticated mapping from the current atmospheric state to the state a
few hours or days hence; the weights used in the neural network are
chosen to minimize error, often root-mean square, mean absolute error, 
or some measure of distributional distance such as the Continuous 
Ranked Probability Score (CRPS; Gneiting and Raftery 2007).
These mappings are typically chained together to provide a prediction;
from the current state, a forecast is made to six hours in the future;
from the forecast at six hours, a forecast is made to twelve hours
hence, and so forth. Henceforth we will refer to these data-driven
models as deep-learning NWP models, or DLNWP. The complexity of these
models is substantial but is hidden within the neural network; the
actual number of lines of code tailored to the prediction application is
very small compared to conventional NWP, perhaps by a factor of 100 or
more.

The first low-resolution, proof-of-concept DLNWP models were developed
only in the late 2010's (Dueben et al. 2018) and were significantly less
accurate than conventional NWP forecasts. Informed by these early test
systems, more computational horsepower was made available to train more
sophisticated DLNWP models, and these have increased in skill at a
dramatic rate. Advanced DLNWP development was demonstrated in Weyn et
al. (2020, 2021), GraphCast (Lam et al. 2023), Met-Net-3 (Andrychowicz
et al. 2023), NeuralGCM (Kochkov et al. 2024), GenCast (Price et al.
2024), FourCastNet (Pathak et al. 2022) and CorrDiff (Mardani et al.
2023), ClimaX (Nguyen et al. 2023), Aurora (Bodnar et al. 2024), FengWu
(Han et al. 2024), FuXi (Chen et al. 2023), AIFS (Lang et al. 2023),
WeatherNext~2 (Alet et al. 2025), FourCastNet3 (Bonev et al. 2025),
HourGlass (Ingstad et al. 2026), STRATA (Hu et al. 2026),
and StormScope (Pathak et al. 2026), and many more. DLNWP
represents a radical change for weather prediction -- simplified code
that bypasses the many complex parameterizations of physical processes
and yet may produce more accurate forecasts. In selected ways of
measuring weather forecast skill, several of these are now competitive
with or more skillful than raw conventional numerical forecasts from the
world-leading European Centre for Medium-Range Weather Forecasts
(ECMWF).

Alongside these global DLNWP models, some progress has recently been made
at regional and convection-permitting scales. StormCast (Pathak et al.
2024) demonstrated a generative diffusion model seeking to emulate
the National Oceanic and Atmospheric Administration 
(NOAA) 3-km High-Resolution Rapid Refresh (HRRR) over the CONUS.
Ba\~{n}o-Medina et al. (2025)
trained a stretched-grid regional AI model at 6-km resolution over the
western United States, achieving performance competitive with operational
regional NWP for 24-h accumulated precipitation and improved skill in
capturing extreme events associated with atmospheric rivers. HRRRCast
(Abdi et al. 2026) further examined data-driven HRRR emulation over the
CONUS. HiRO-ACE (Perkins et al. 2025) trained a
stochastic diffusion-based downscaling framework on a decade of global
3-km storm-resolving simulations. Limited-area
model (LAM) architectures for regional DLNWP have also advanced
quickly: Larsson et al. (2025) introduced Diffusion-LAM, which uses
conditional diffusion with lateral boundary forcing to produce
probabilistic regional forecasts, while Adamov et al. (2025)
systematically evaluated design choices for km-scale ML LAMs across
European domains with varied terrain. Sha et al. (2026) applied an AI-based
regional downscaling system over the southern Great Plains and
southeastern United States, demonstrating stable performance across
multi-decadal simulations including future climate scenarios. Despite
this rapid progress, the scarcity of high-resolution training data at
convection-permitting scales remains a key constraint on the further
development of regional DLNWP systems.

Training data are needed for such projects.    What training data would be ideal?
A long time series (say, a decade or more) and a large domain are certainly desirable, 
thus spanning a wide range of atmospheric phenomena and landscapes
so that the trained model can generalize well.
Despite the extra expense of storage, high resolution is desirable, ideally of
order 1 km so that individual thunderstorms are better resolved. The ideal 
training data would also have a high temporal cadence, 15 min or less,
so that the data are available to model the life cycle of phenomena like short-lived
thunderstorms.   Further, the data would have the coherent
temporal evolution of a classical numerical weather prediction system but 
would have minimal biases such as are characterized by reanalyses 
such as in Hersbach et al. (2020) and Hamill et al. (2021).
Unfortunately, direct high-resolution regional reanalysis production would be very 
computationally demanding, and a direct regional reanalysis may not exhibit 
the temporal continuity that is beneficial to DLNWP training.   
An appealing approach may be a global reanalysis nudging 
to a high-resolution model simulation.   This approach was used in ``CONUS404'', 
Rasmussen et al. (2023).  A disadvantage of the particular CONUS404 data set, 
however, is that most variables were not saved at sub-hourly time steps
that facilitate thunderstorm prediction.  

\begin{figure}[htbp]
\centering
\includegraphics[width=6.5in]{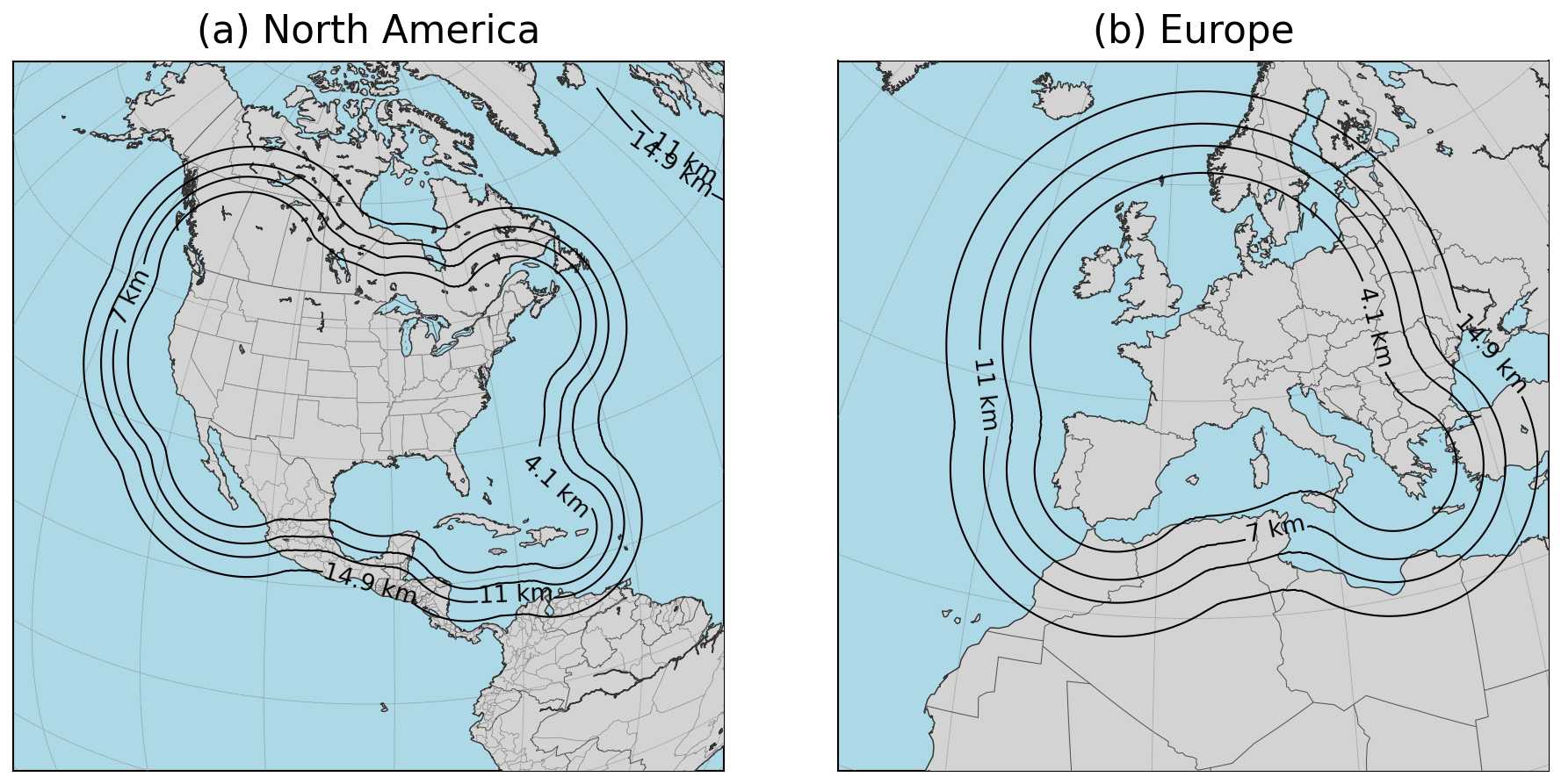}
\caption{Illustration of the variable grid spacing for the
global MPAS model to be used in the reforecasts described here. Grid
spacing is \textasciitilde4 km inside the innermost line, \textasciitilde15 km
outside the outermost.  Used with permission from Hamill (2026).}
\label{fig:1}
\end{figure}

Given computational and time constraints, when seeking to develop a
DLNWP emulator of thunderstorm-scale weather, we chose to generate a reforecast rather
than a reanalysis.  A trained emulator based on these data would have the disadvantage of 
producing guidance that emulates a forecast rather than the real atmosphere,
so statistical postprocessing is likely necessary to ameliorate bias before
product dissemination. To jumpstart our own convection-permitting DLNWP, we 
generated a convection-permitting reforecast data set with mesh refinement 
to 4 km over much of North America and the Caribbean, as well as western Europe
(Fig.~\ref{fig:1}).   This is the configuration of the operational 
``GRAF'' (Global high-Resolution Atmospheric Forecasting) model run at
The Weather Company, described in the next section. Most data are saved at 
15-min cadence, some with 5-min cadence. Our companies, NVIDIA 
and The Weather Company, have subsequently developed DLNWP methods 
with these data. We now provide the data without cost to enterprise partners 
to facilitate others' ability to advance their own DLNWP.

The rest of the document describes model configuration details, the
procedure that was used to define the dates of initial conditions, and
information on the data set storage.

\section{Model configuration}

The TWC GRAF model
is based on the NCAR MPAS (Model for Prediction Across Scales; Skamarock
et al. (2012), Park et al. (2014), Sandbach et al. (2015), Heinzeller et
al. (2016)), version 6.3. Basic details are included in Table~\ref{tab:1}, and
zeta values used in the determination of 50 model-level heights are
provided in Table~\ref{tab:2}. The model has a \textasciitilde{} 4-km grid spacing
over the CONUS and Europe, relaxing to 15 km elsewhere (Fig.~\ref{fig:1}). It
employs a scale-aware nTiedtke convective parameterization (Wang 2022)
with near full use of the convective parameterization at 15 km to near
none at 4 km. It also uses the Yonsei University (YSU) planetary boundary layer and
gravity-wave drag, WRF Single-Moment 6-class (WSM6) microphysics (Hong and Lim 2006) with Thompson
cloud fraction, Rapid Radiative Transfer Model for GCMs (RRTMG) radiative transfer scheme (Mlawer et al. 1997,
Clough et al. 2005), and the NOAH 4-layer land-surface model (Ek et al.
2003). Terrain data are from ``GMTED2010/MODIS 30 arcsec'' (Global
Multi-resolution Terrain Elevation Data
(\href{https://www.usgs.gov/coastal-changes-and-impacts/gmted2010}
{\underline{https://www.usgs.gov/coastal-changes-and-impacts/gmted2010}})),
and albedo and land characteristic data are from the Moderate Resolution Imaging Spectroradiometer (MODIS) 30 arcsec
(\href{https://modis-land.gsfc.nasa.gov/brdf.html}
{\underline{https://modis-land.gsfc.nasa.gov/brdf.html}}).
The model was initialized from ERA5 reanalyses on model levels (Hersbach
et al. 2020).  These data do not have the high-resolution information of 
our operational GRAF data assimilation, which assimilates recent 
observations using 3-dimensional variational analysis 
cold-started from the most recent operational ECMWF
analysis.

\begin{longtable}[]{@{}p{1.7in}p{4.6in}@{}}
\caption{Configuration for the GRAF reforecast.}\label{tab:1}\\
\toprule
\textbf{Model} & MPAS 6.3 (using GPUs, with TWC tuning), 50 levels, \textasciitilde{} 4.8 M grid cells \\
\midrule
\endhead
\textbf{Physics} & WSM6 cloud microphysics; nTiedtke scale-aware convective parameterization; YSU planetary boundary layer and gravity wave drag; surface layer via Monin-Obukhov option. RRTMG radiative transfer scheme; NOAH land-surface model; Thompson cloud fraction \\
\textbf{Initialization} & MPAS 7.0 \\
\textbf{Terrain} & Global Multi-resolution Terrain Elevation Data 2010 \\
\textbf{Albedo} & MODIS Land, 30 arcsec grid spacing. \\
\textbf{Land Use} & MODIS-derived predominant land type, veg fraction. \\
\textbf{Atmospheric state, sea-surface temperature, soil state} & ERA5 reanalysis on model levels, 0.25-deg. \\
\bottomrule
\end{longtable}

\begin{longtable}[]{@{}llllll@{}}
\caption{A list of the staggered zeta values used in conjunction with
terrain elevation to determine the 50 interpolated model coordinate heights
following Klemp (2011).}\label{tab:2}\\
\toprule
\textbf{Model level} & \textbf{Zeta (m)} & \textbf{Model level} &
\textbf{Zeta (m)} & \textbf{Model level} & \textbf{Zeta (m)} \\
\midrule
\endfirsthead
\multicolumn{6}{l}{\textit{Table~\thetable~(continued)}}\\[2pt]
\toprule
\textbf{Model level} & \textbf{Zeta (m)} & \textbf{Model level} &
\textbf{Zeta (m)} & \textbf{Model level} & \textbf{Zeta (m)} \\
\midrule
\endhead
1 & 0 & 18 & 7528 & 35 & 24256 \\
2 & 66 & 19 & 8512 & 36 & 25240 \\
3 & 150 & 20 & 9496 & 37 & 26224 \\
4 & 250 & 21 & 10480 & 38 & 27208 \\
5 & 380 & 22 & 11464 & 39 & 28192 \\
6 & 550 & 23 & 12448 & 40 & 29176 \\
7 & 750 & 24 & 13432 & 41 & 30160 \\
8 & 1000 & 25 & 14416 & 42 & 31144 \\
9 & 1300 & 26 & 15400 & 43 & 32128 \\
10 & 1700 & 27 & 16384 & 44 & 33112 \\
11 & 2100 & 28 & 17368 & 45 & 34096 \\
12 & 2600 & 29 & 18352 & 46 & 35080 \\
13 & 3200 & 30 & 19336 & 47 & 36064 \\
14 & 3900 & 31 & 20320 & 48 & 37048 \\
15 & 4700 & 32 & 21304 & 49 & 38032 \\
16 & 5600 & 33 & 22288 & 50 & 39016 \\
17 & 6544 & 34 & 23272 & 51 & 40000 \\
\bottomrule
\end{longtable}

\section{Approach for choosing the dates of initial conditions}

The initial-condition dates were selected in multiple ways but
generally prioritized dates expected to have high-impact weather in the
contiguous US. This approach was motivated by recent research on weather
importance sampling,  including Kravtsov et al. (2022) and Lancelin et al. (2026).
These included choosing cases for near-landfalling
US/Mexican hurricanes, US severe local storms, and dates with previously
forecast heavy precipitation in major US hydrologic units. A small
number of cases were selected at the end of the process, filling in the
largest gaps in sequences of dates for weather-dependent reforecasts.

\subsection{Selecting the dates for tropical cyclones, 
severe local storms, and other high-impact weather phenomena}

This part of the initial-condition date selection process was not fully
objective in character. Wikipedia pages for US Atlantic hurricane
seasons from 2004--2023 were examined (e.g.,
\href{https://en.wikipedia.org/wiki/2004_Atlantic_hurricane_season}{\underline{here}}),
and initial dates were chosen typically 12--24 h preceding US or near-US
land-falling hurricanes. Occasionally cases were chosen for US tropical
storms as well. With hurricanes that lingered over land such as Harvey
in 2017 (Houston-area floods), or with hurricanes that stayed near the
coast such as 2019's Dorian, multiple reforecast dates were chosen
closely spaced in time and covered more than just landfall. The only
eastern Pacific hurricane date was for Otis (2023), which rapidly
intensified unexpectedly and made landfall at Acapulco, Mexico. Given
the reforecasts extend to only a short lead time, we do not envision
this method as providing, in itself, a quantification of track or
intensity skill beyond the 27 h of the reforecast. Hurricane Sandy
(2012) was accidentally omitted.

Many case dates were also selected based on the criteria of either
observed or forecast severe weather and tornado outbreaks. Again,
Wikipedia pages were examined for severe local storms (e.g.,
\href{https://en.wikipedia.org/wiki/Tornadoes_of_2004}{\underline{here}}),
and dates with many tornadoes or a few significant tornadoes were
chosen. Since most tornadoes occur in late afternoon, typically a 12 UTC
initialization time was commonly chosen. A few cases with overnight
tornadoes were initialized at 18 or 00 UTC.

We also examined the literature for dates of predecessor rain events
(PREs, Galarneau et al. 2010) associated with tropical cyclones, major
northeast US snowstorms, ice storms, and major west-coast atmospheric
rivers (Zhu and Newell 1994), and we included a few dates for these.
These lists are admittedly incomplete; the list of PREs was based on
dates listed in a 2010 journal article, so all cases precede 2010.
Northeast US snowstorms were based on Google searches and the memories
of TWC employees. We wanted to find dates of major mesoscale convective
complexes, as these are often poorly modeled in numerical guidance and
occur in weaker synoptic flow regimes; we found no online references for
these, regrettably.

\begin{figure}[htbp]
\centering
\includegraphics[width=6.5in,height=6.5in]{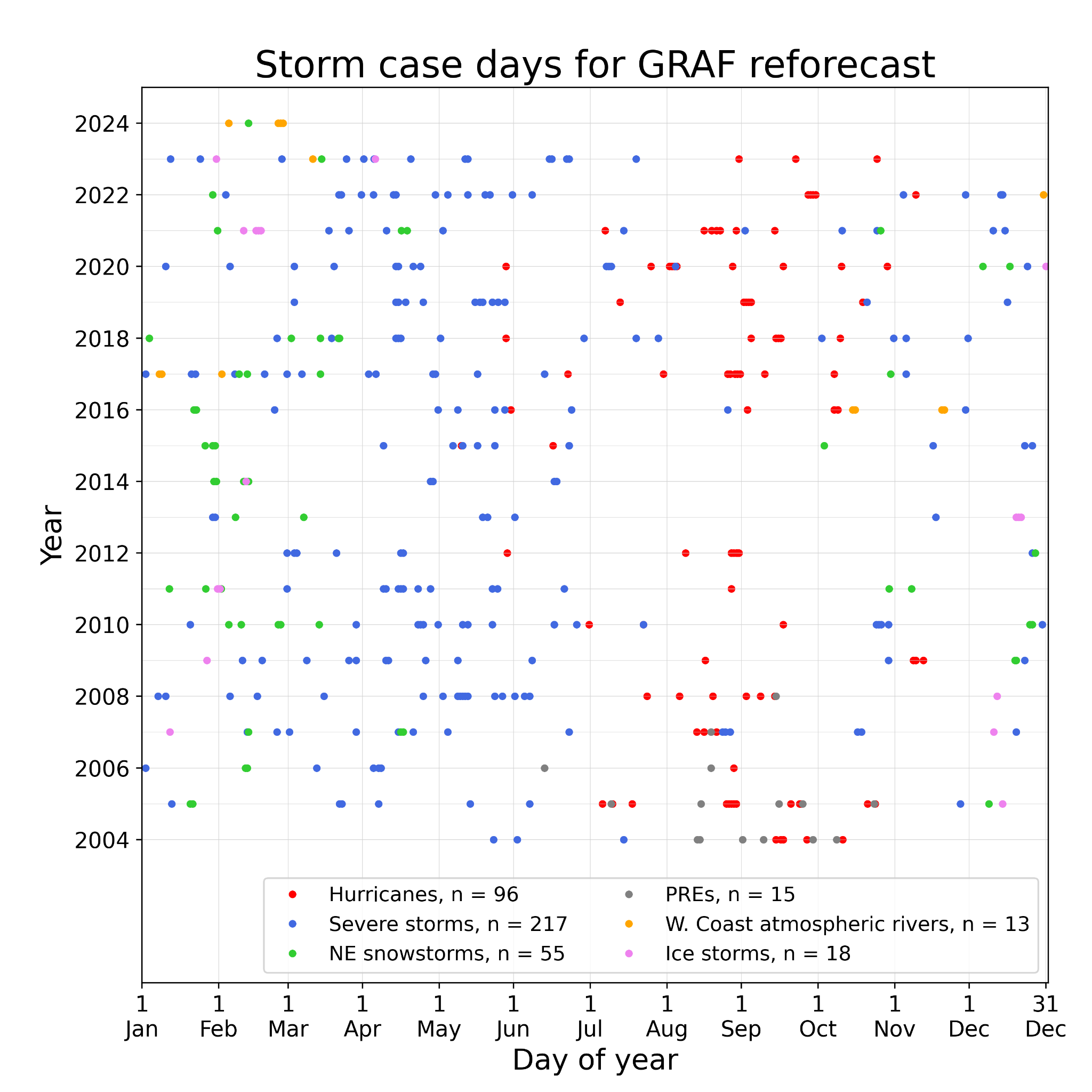}
\caption{Illustration of the 396 dates chosen for severe
local storms, hurricanes, northeast US snowstorms, predecessor rain
events, and west-coast atmospheric rivers.}
\label{fig:2}
\end{figure}

A synthesis of the cases selected based on these criteria is shown in
Fig.~\ref{fig:2}.

\subsection{Selecting most of the cases based on heavy forecast rainfall in major hydrologic units}

Since the primary initial use case for the TWC-NVIDIA collaboration is
to develop and demonstrate a capacity for high-resolution, deep-learning
based ensemble forecasts of precipitation, most of the remaining cases
were selected based on heavy forecast ensemble-mean precipitation
somewhere in the CONUS. This provides us with a sample that has more
heavy precipitation forecasts than normal, and presuming a positive
forecast-to-observation correlation, more heavy observed forecasts than
normal. Will this bias the machine learning training? Maybe so, but if
we wish to be able to train the deep learning algorithm to provide
reasonable results spanning the most dry to the most wet scenarios, we
will need sufficient samples of both wet and dry forecasts, ideally with
wet samples across the domain (Lancelin et al. 2026).
The assumption underlying the selection
of a subset of wet forecast case dates for each major river basin is
that those case dates will often have dry forecasts in regions outside
that river basin. Thus, as we describe here, partitioning the CONUS into
18 separate river basins and one CONUS-wide area and finding wet cases
independently for each will still result in a wide range of
precipitation scenarios across the set of reforecast cases; a wet
reforecast case day for a western US river basin may be a very dry day
for an eastern US river basin.

To ensure we have heavy precipitation events represented across the US,
we chose a subset of heavy precipitation cases for each major CONUS
hydrologic unit, shown in Fig.~\ref{fig:3} below. These ``HUC-2'' (Hydrologic Unit Code) units were
defined by the US Geological Survey (Seaber et al. 1987). Additionally,
we considered precipitation averaged over all HUC-2 units.

\begin{figure}[htbp]
\centering
\includegraphics[width=6.5in,height=5.19444in]{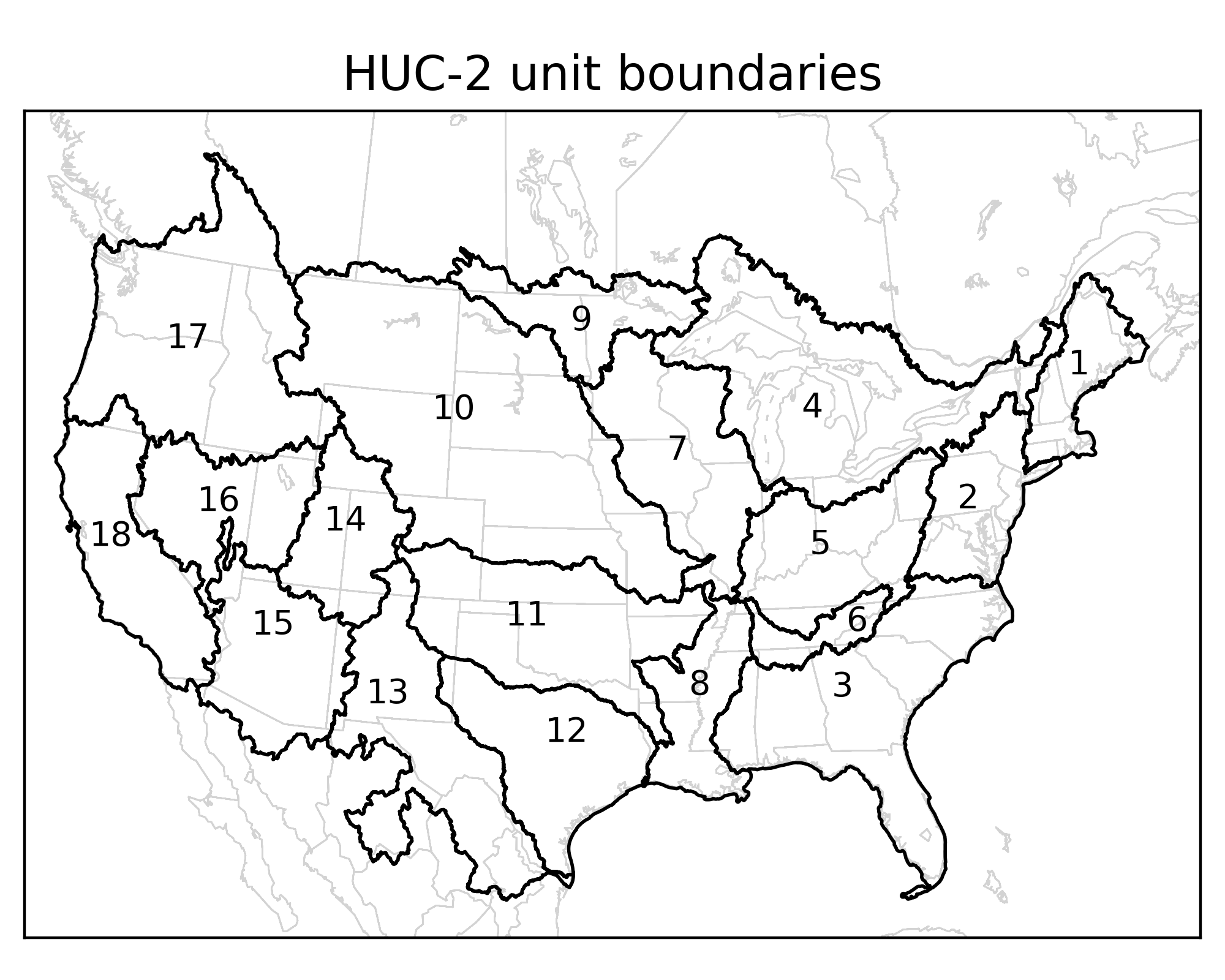}
\caption{Illustration of HUC-2 units for the CONUS for which
heavy precipitation cases were chosen. There are 18 of these, identified
as numbered. The methodology also chose cases based on a 19th area, all
18 HUCs together.}
\label{fig:3}
\end{figure}

We confined the reforecasts initial conditions to be primarily in the
20-year period of 2004--2023, with a few cases in early 2024; choosing
initial condition dates further into the past has the disadvantage of
being initialized with data from a thinner observation network;
forecasts from the initial conditions in 1990 would thus have a lower
statistical quality and higher errors than forecasts from initial
conditions in recent decades. There is also the complication of the
changing climate, with more extremes of precipitation in recent decades.

We chose initially to generate 715 HUC-2 case dates for both the cool
season (October--March) and warm season (April--September). For each
season, we divided those cases with 35 for each HUC-2 unit and the
remainder, 85 samples, for the overall CONUS.

The algorithm for selecting those case dates is as follows, repeated
independently for each HUC-2 unit. First, in order to avoid biasing the
cases too much toward heavy \emph{observed} precipitation, these cases
are selected based on heavy \emph{forecast} precipitation. Global Ensemble Forecast System (GEFS) v12
total-precipitation reforecasts for 2004--2019 (Guan et al. 2022) were
used as the input data for case selection in these years. All the GEFS
reforecasts were generated from 00 UTC initial conditions. GEFS v12
forecasts and reforecasts have a notable issue with spin-up, i.e., a
change in the statistical character of the forecast in the forecasts'
early hours. In the case of GEFS v12, there is typically anomalously
heavy precipitation in the first 12 hours. Hence, as a way of
quantifying forecast precipitation intensity, we determined, spatially
averaged over each HUC-2 region, the 12--36 h ensemble-mean
precipitation, a 24-h period after the worst of the spin-up should have
taken place.

GEFS v12 reforecasts extend to the end of 2019, while the reforecast
cases were desired for 2020--2024 as well. For these other years, TWC
used internal data. We maintain a database of GEFS ensemble forecasts
coincident with regularly reporting observation stations. For similarity
to the procedure above, we determined the 12--36-h ensemble-mean
forecast, averaged over all the stations within each HUC-2 unit.

In general, for most HUC-2 units, there are fewer stations for the
2020--2023 data than there were model grid points for the 2004--2019
reforecasts. Hence, from central limit theorem arguments, we would
expect a higher variance from the 2020--2023 station-based data. To make
the statistics more uniform across the full 2004--2023 period before
selecting case dates, we standardize the 2020--2023 data and then
re-express with the mean and variance of the 2004--2019 data. As we
process either the warm or cool season, let \({\mu}_{r}\) represent the
mean of all that season's daily 2004--2019 reforecast samples for a
HUC-2 unit, and let \(\sigma_{r}\) be the climatological standard
deviation of those daily samples. Similarly, we compute
\({\mu}_{s}\) and \(\sigma_{s}\), the mean and standard deviation based
on 2020--2023 GEFS ensemble-mean forecasts at observation locations.
Letting \(x_{i}\) be the GEFS station-based ensemble-mean forecast for
the $i$th case day in the 2020--2023 period,
\({\widehat{x}}_{i}\) is the re-expressed mean, computed as

\begin{equation}
\widehat{x}_i = {\mu}_{r} + \sigma_{r}\frac{x_{i} - {\mu}_{s}}{\sigma_{s}}
\end{equation}

For 2020--2023, these re-expressed means then replace the raw ensemble-
and station-mean averages in the time series for each basin.

\begin{figure}[htbp]
\centering
\includegraphics[width=6.5in,height=5.31944in]{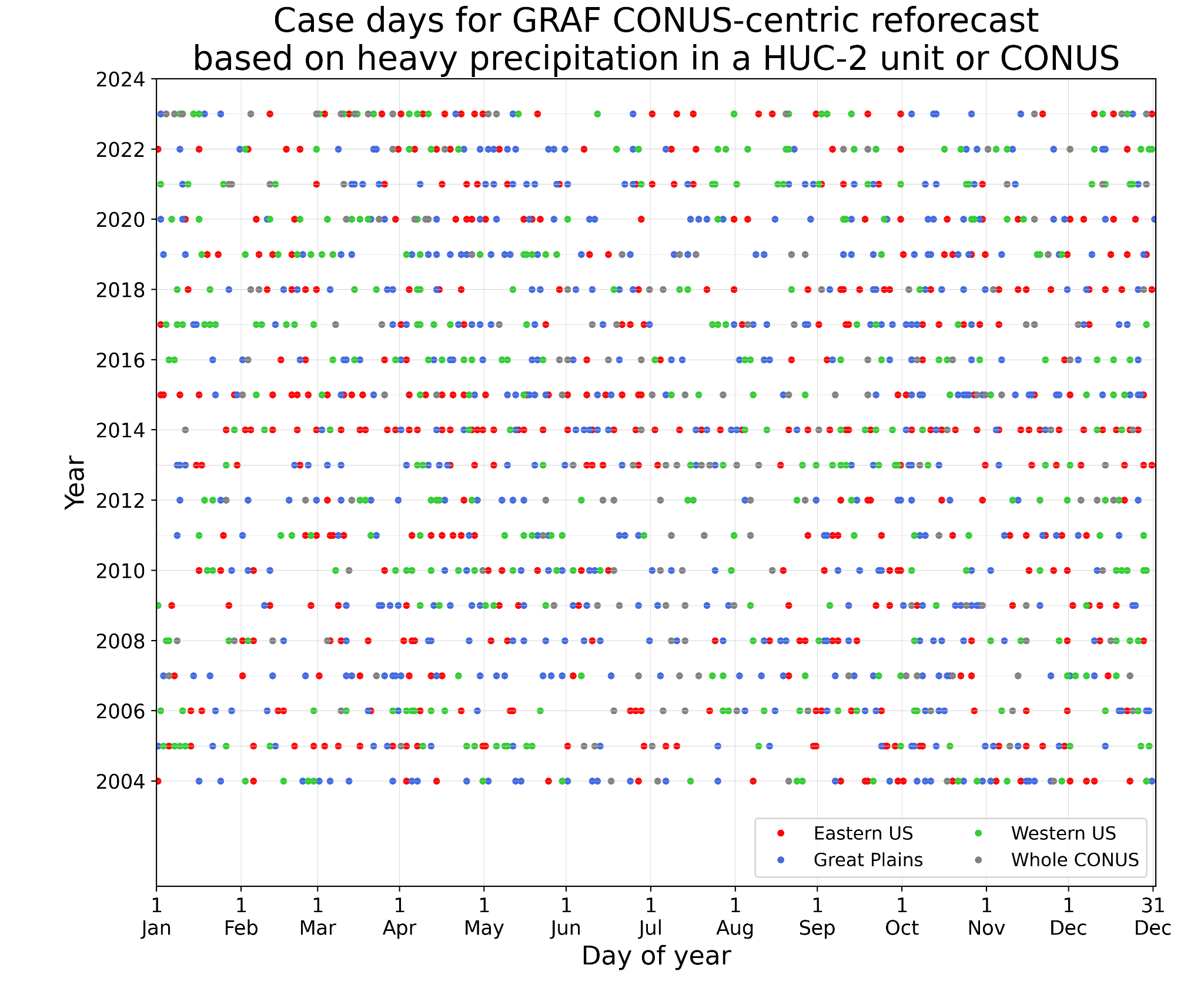}
\caption{Case dates selected based solely on the GEFS 12--36h
precipitation. These are shown for three different regions, eastern US
(HUC-2 units 1--6), Great Plains (units 7--13), and western US (units
14--18), in addition to the whole CONUS.}
\label{fig:4}
\end{figure}

The cases were then selected based largely on a rank ordering of
ensemble-mean precipitation. Case dates were ordered from lowest to
highest ensemble-mean precipitation, and they were assigned an initial
weight based on that date's ensemble-mean precipitation, divided by the
ensemble-mean precipitation for the date with the largest value. The
first case date selected was the date with the largest precipitation,
i.e., with the largest weight. Then, with one exception, the second and
third dates have the second and third largest precipitation, and so
forth through the samples with progressively less precipitation. The one
exception is that in order to slightly de-emphasize the selection of
dates that may be very close in time so that we have more independent
samples, if a particular initial condition date was selected, the nearby
dates were de-weighted so they were less likely to be chosen. For the
day before, two days before, the day after, and two days after, the
initial weights as described above were multiplied by a factor of 0.7.
This would thus not totally eliminate such dates from consideration but
gave them less probability of being chosen. The weight factor value of
0.7 was somewhat arbitrary; the rationale was that should a very major
precipitation event span multiple days, we would still want a higher
probability of choosing it. Figure~\ref{fig:4} shows the HUC-2 case dates selected
with the process described above.

\begin{figure}[htbp]
\centering
\includegraphics[width=6.5in,height=5.31944in]{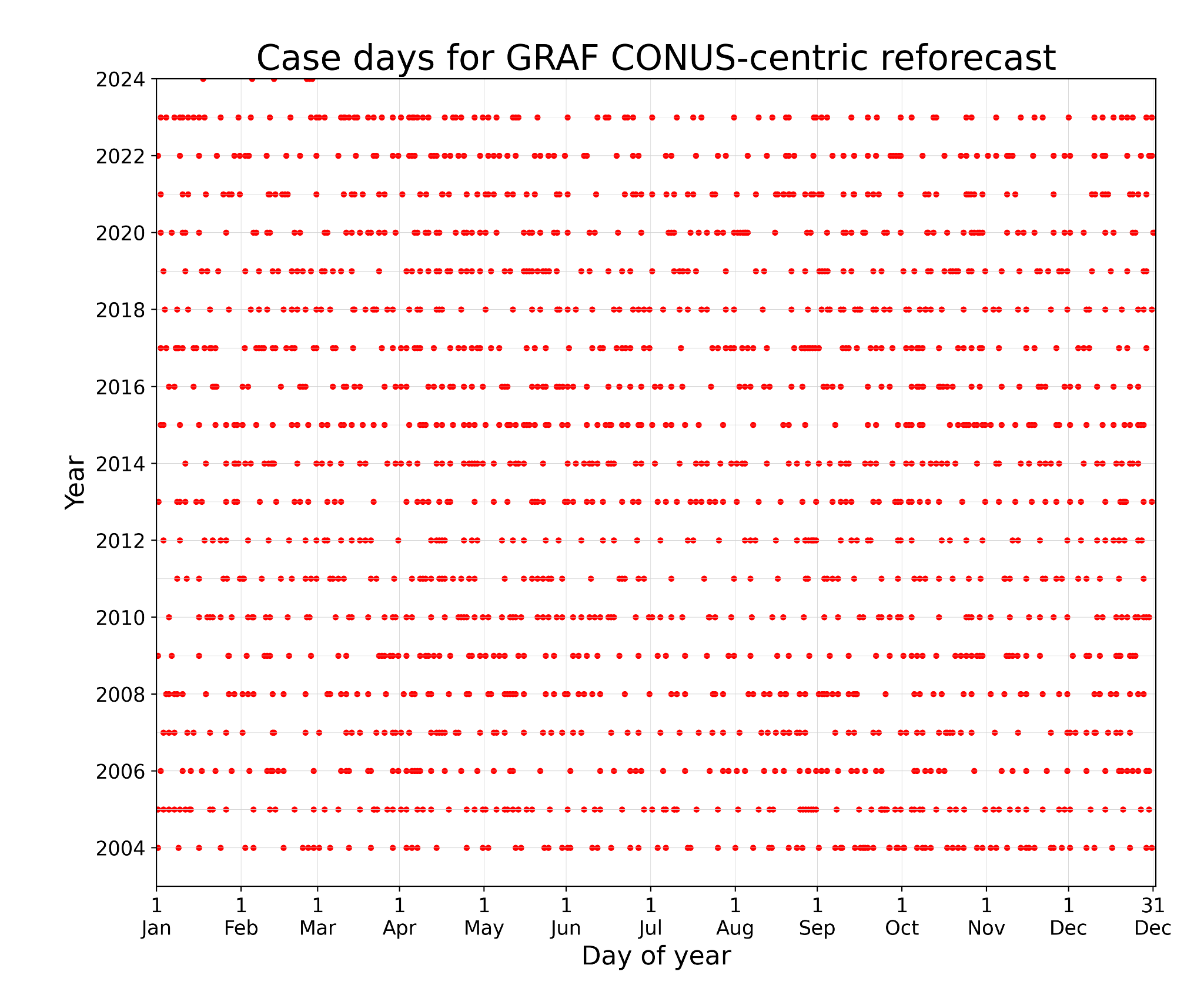}
\caption{The final proposed list of GRAF reforecast case
dates, including cases selected for major storms of varying types, heavy
precipitation, and gap filling.}
\label{fig:5}
\end{figure}

At this point, there was no combination yet with the other cases chosen
for hurricanes, severe local storms, snowstorms, and such. When we
merged these data, there were some overlapping initial-condition dates,
perhaps for a date chosen for a land-falling hurricane that was
simultaneously chosen for the heaviest precipitation in that basin. We
eliminated the overlaps, freeing up a small number for new cases to make
the grand total of 1836 cases. We chose to use these cases to fill in
the largest gaps in dates. Suppose the longest gap between cases was 20
days; then the first infill initial condition case date is the one in
the middle of that gap. We then proceed to the next-longest gap, filling
that, and so forth. The final list of cases is shown in Fig.~\ref{fig:5}. The
distribution is not random. Some periods, such as the last half of 2004,
were exceptionally stormy and had more than the average number of case
dates. In comparison, the last half of 2023 had far fewer cases. This
strategy, we hope, will allow us to train deep learning models focused
on precipitation and achieve acceptable accuracy with a smaller number
of cases than might be necessary with a regular sampling strategy. Such
a hypothesis is difficult to test, for this would require reforecasts
for both sampling strategies.

\section{Stored data from the reforecast}

Table~\ref{tab:3} shows the variables that were saved, their temporal resolution,
and whether they were 2-D fields or 3-D. These data were stored in zarr
format at \texttt{s3://twc-graf-reforecast/}. The data are not spatially
chunked. There are separate zarr files for the variables archived with
5-minute output data and 15-min output data. 

Data have been compressed.
To keep the total storage volume reasonable, data were compressed using 
an error-bounded lossy compression scheme which included:
\begin{itemize}
    \item An optional Normalize filter that performs z-score normalization
        of variables using a fixed prescribed mean and standard deviation.
    \item BitRound (Klöwer et al. 2021) method truncating to 10 mantissa bits.
    \item A byte shuffle filter to split 32 bit floating point elements into
        4 independent byte streams. This allows the downstream compression
        algorithm to leverage patterns in the exponent bits or the most
        significant mantissa bits allowing for more efficient compression.
    \item Zlib compression using DEFLATE level 9 for maximum compression
        ratio.
\end{itemize}
The rationale for choosing this compression pipeline was to match the
error-bounded lossy compression specification to what would be meaningful
to downstream training pipelines and discard bits that do not significantly
contribute to model training. Since most neural networks are trained in
reduced precision formats like Tensor-Float32, IEEE Float16 or Bfloat16,
keeping all 23 mantissa bits of IEEE float formats is unnecessary and we
can truncate to 10 mantissa bits (the max of the three aforementioned 
formats). This is essentially what the BitRound filter does with the 
error-bounds being relative absolute error bounds. One caveat of
using BitRound with variables like temperature or pressure is that they
have large means compared to their standard deviation and a relative 
error bound is not appropriate here. To tackle this issue, we first 
perform z-score normalization of each variable to remove the global mean
before using BitRound as a way to preserve the maximum fidelity within 
each variable's dynamic range given a fixed bit budget. For variables where
the total domain integral is important or where it is important to preserve
zero values, like precipitation variables, we specifically set the mean 
used in the Normalize filter to zero. This prevents bit truncation related
zero-drift issues as well as preserves total domain integrals for
precipitation variables. Using this compression pipeline, we are able
to compress the dataset by a factor of \textasciitilde6 which makes
storage and distribution of the dataset feasible.

\begin{figure}[htbp]
\centering
\includegraphics[width=6.5in,height=6.31944in]{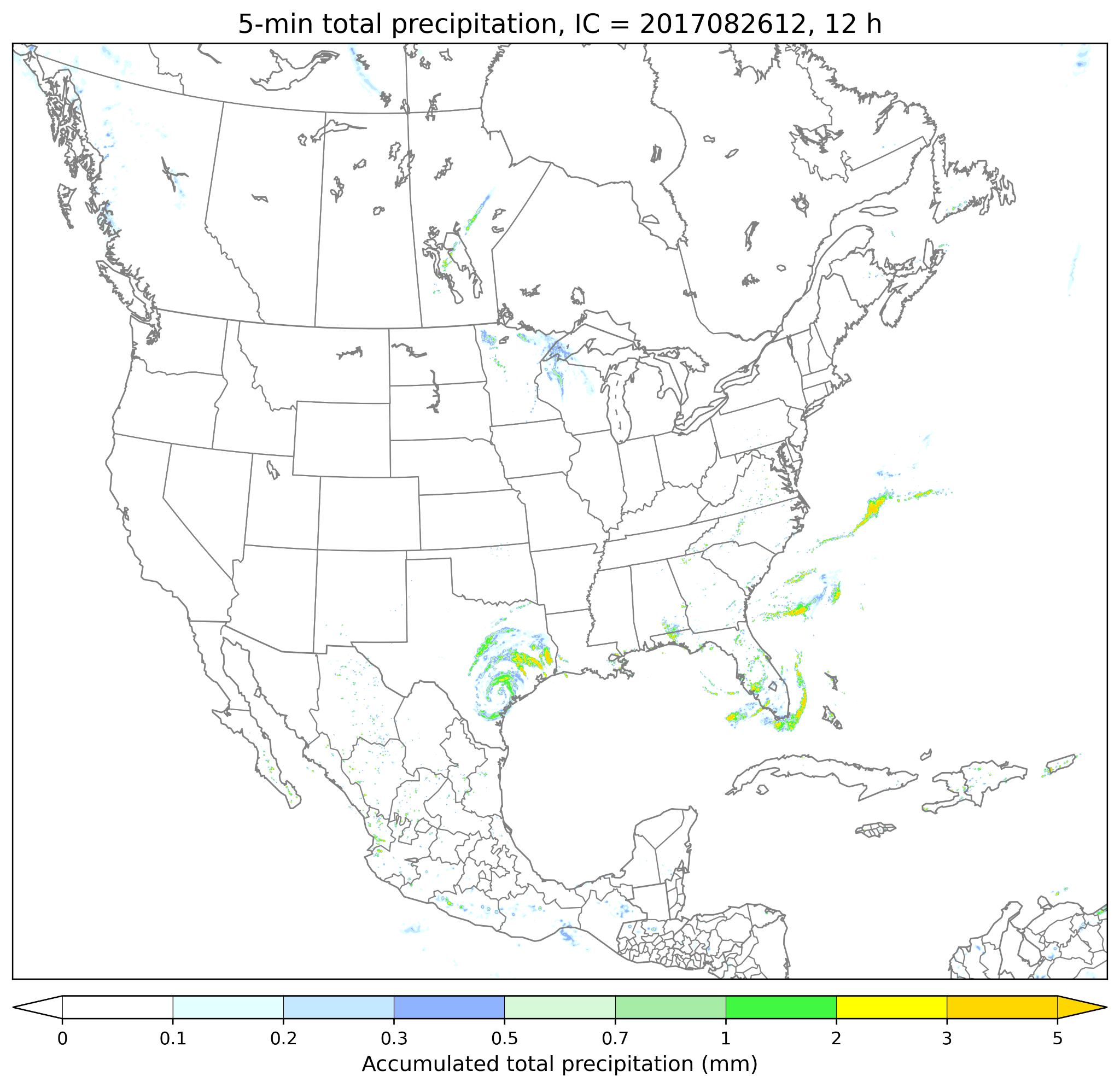}
\caption{An illustration of 5-min precipitation amount shortly
after the landfall of hurricane Harvey.}
\label{fig:6}
\end{figure}

A demonstration python script for reading single-level model data and
producing plots is included in the documentation. An example plot of
model output is shown in Fig.~\ref{fig:6}.

Please note these data are provided on the native model vertical
coordinate. The user will need to perform for themselves a remapping to
other coordinates such as a vertical pressure coordinate.  Software
for performing this is available at \href{https://mpas-dev.github.io/MPAS-Tools/master/index.html}{\underline{https://mpas-dev.github.io/MPAS-Tools/master/index.html}}.

\begingroup
\footnotesize
\begin{longtable}[]{@{}>{\hangindent=1em\hangafter=1\relax}p{1.75in}p{1.8in}p{0.8in}p{0.65in}p{0.75in}@{}}
\caption{List of forecast output variables for the GRAF
reforecast, their temporal frequency, and whether the data are 2-D or
3-D fields. ``Atm'' indicates atmospheric levels; ``soil'' indicates soil
levels.  Items marked with * are not standard MPAS output but rather
are derived specially within The Weather Company.}\label{tab:3}\\
\toprule
\textbf{Variable} & \textbf{Variable name(s) in zarr files} &
\textbf{Temporal frequency} & \textbf{2D or 3D} & \textbf{Units} \\
\midrule
\endfirsthead
\multicolumn{5}{l}{\textit{Table~\thetable~(continued)}}\\[2pt]
\toprule
\textbf{Variable} & \textbf{Variable name(s) in zarr files} &
\textbf{Temporal frequency} & \textbf{2D or 3D} & \textbf{Units} \\
\midrule
\endhead
Water vapor mixing ratio & qv & 15 min & 3D (atm) & kg/kg \\
Cloud water mixing ratio & qc & 15 min & 3D (atm) & kg/kg \\
Rain water mixing ratio & qr & 15 min & 3D (atm) & kg/kg \\
Ice mixing ratio & qi & 15 min & 3D (atm) & kg/kg \\
Snow mixing ratio & qs & 15 min & 3D (atm) & kg/kg \\
Graupel mixing ratio & qg & 15 min & 3D (atm) & kg/kg \\
u-wind component & uReconstructZonal & 15 min & 3D (atm) & m/s \\
v-wind component & ureconstructMeridional & 15 min & 3D (atm) & m/s \\
w-wind component & w & 15 min & 3D (atm) & m/s \\
Total pressure & p & 15 min & 3D (atm) & Pa \\
Temperature & temperature & 15 min & 3D (atm) & K \\
Potential temperature & theta & 15 min & 3D (atm) & K \\
Soil temperature & tsl & 15 min & 3D (soil) & K \\
Soil equivalent liquid water & sh2o & 15 min & 3D (soil) &
m\textsuperscript{3}/m\textsuperscript{3} \\
Soil moisture & smois & 15 min & 3D (soil) &
m\textsuperscript{3}/m\textsuperscript{3} \\
Kuchera snow ratio & snow\_ratio & 15 min & 2D & m/m \\
Total snow depth & snowh & 15 min & 2D & m \\
Visibility & visibility & 15 min & 2D & m \\
Conditional probability of rain & cporain & 15 min & 2D & n/a \\
Conditional probability of snow & cposnow & 15 min & 2D & n/a \\
Conditional probability of ice & cpoice & 15 min & 2D & n/a \\
10-m u component & u10 & 15 min & 2D & m/s \\
10-m v component & v10 & 15 min & 2D & m/s \\
All-sky downward surface shortwave radiation flux & swdnb & 15 min & 2D
& W/m\textsuperscript{2} \\
Downward surface shortwave Direct Normal Flux & swdnbdn & 15 min & 2D &
W/m\textsuperscript{2} \\
Downward all-sky surface flux, short and longwave, 1-h average &
swdnb01h & 15 min & 2D & W/m\textsuperscript{2} \\
Downward surface shortwave Direct Normal Flux, 1-h average & swdnbdn01h
& 15 min & 2D & W/m\textsuperscript{2} \\
Precipitation rate & prate & 5 min & 2D & mm/s \\
Predominant precipitation type & ptype & 5 min & 2D & n/a \\
Rain accumulation & rain\_bucket & 5 min & 2D & mm \\
Convective rain accumulation & convective\_bucket & 5 min & 2D & mm \\
Ice accumulation & zrain\_bucket & 5 min & 2D & mm \\
Snow accumulation & snow\_bucket & 5 min & 2D & mm \\
Total precipitation accumulation & apcp\_bucket & 5 min & 2D & mm \\
Total cloud cover & total\_cloud\_cover & 5 min & 2D & \% \\
Mean sea-level pressure & mslp & 15 min & 2D & Pa \\
2-m temperature & t2m & 15 min & 2D & K \\
2-m dewpoint & dewpoint\_2m & 15 min & 2D & K \\
2-m specific humidity & q2 & 15 min & 2D & kg/kg \\
Total-column precipitable water & precipw & 15 min & 2D &
kg/m\textsuperscript{2} \\
Skin temperature, including SST & skintemp & 15 min & 2D & K \\
Wind gust & windgust10m & 15 min & 2D & m/s \\
All-sky top of atmosphere outgoing longwave & olrtoa & 15 min & 2D &
W/m\textsuperscript{2} \\
Lifted index & bli & 15 min & 2D & K \\
Convective available potential energy & cape & 15 min & 2D & J/kg \\
Convective inhibition & cin & 15 min & 2D & J/kg \\
Lifted condensation level & lcl & 15 min & 2D & m \\
Ceiling above ground level & ceiling\_agl & 15 min & 2D & m \\
Echo top (18 dBz)* & echotop & 15 min & 2D & m \\
Fire weather index* & fwi & 15 min & 2D & n/a \\
PBL height & hpbl & 15 min & 2D & m \\
Latent heat at the surface & lh & 15 min & 2D &
W/m\textsuperscript{2} \\
Hourly averaged latent heat flux & lh01h & 15 min & 2D &
W/m\textsuperscript{2} \\
Power disruption index* & pdi & 15 min & 2D & n/a \\
Thunderstorm potential index* & tpi & 15 min & 2D & n/a \\
\bottomrule
\end{longtable}
\endgroup

\section{Known issues with the reforecast}

\begin{figure}[htbp]
\centering
\includegraphics[width=6.5in,height=2.56944in]{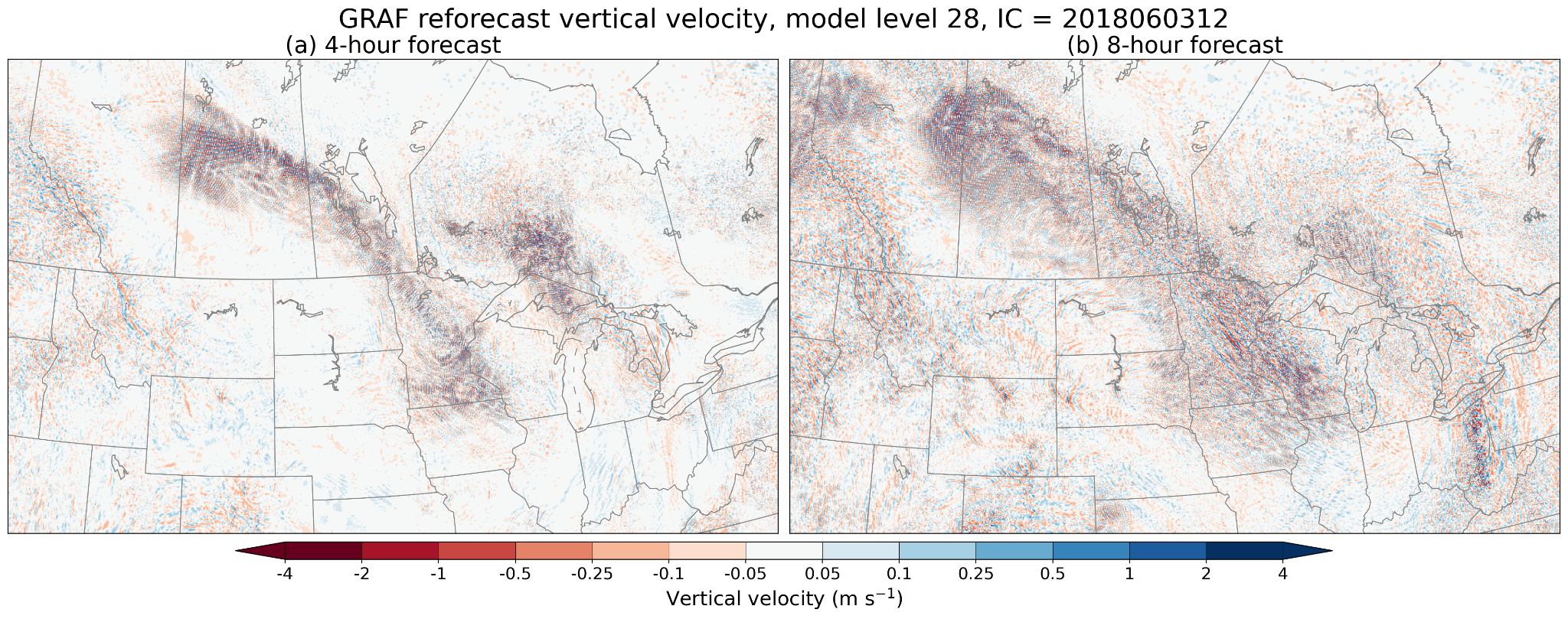}
\caption{An illustration of the growth of anomalous
small-scale vertical velocities during a reforecast, here for a
simulation initialized at 12 UTC on 3 June 2018.}
\label{fig:7}
\end{figure}

A significant issue was discovered during the process of training a GRAF
emulator, and this problem is illustrated in Fig.~\ref{fig:7}. Simulations are
typically contaminated as the simulation progresses by very small-scale
vertical velocities, usually starting small in spatial extent and near
the tropopause, then spreading in horizontal and
vertical extent. If one examines other variables such
as horizontal winds, some minor manifestation of the issue does
become apparent in these, but for the most part, other variables such as
winds and temperature look more reasonable, and surface precipitation
does not show obvious manifestations of the problem in most
circumstances. In our own training of a GRAF emulator, to be described
in a subsequent manuscript, we have found that realistic emulation of
GRAF is still possible by omitting the use of vertical velocity as a
feature and through scale-dependent filtering.

\section*{Acknowledgments}

Some introductory text was borrowed from the NOAA Science Advisory
Board, Environmental Information System Working Group's statement on
deep learning numerical weather prediction (NOAA Environmental
Information Systems Working Group, 2024, available
\href{https://sab.noaa.gov/wp-content/uploads/EISWG-Statement-on-Deep-Learning-NWP_Final_06-18-2024.pdf}{\underline{here}}).
The lead author of this document was the lead author of that statement.

\newpage
\section*{References}

\begin{hangparas}{2em}{1}

Abdi, D., I. Jankov, P. Madden, V. Vargas, T. A. Smith, S. Frolov,
M. Flora, and C. Potvin, 2026: HRRRCast: a data-driven emulator for
regional weather forecasting at convection-allowing scales. \emph{Artif.
Intell. Earth Syst.}, \textbf{5}, e250061.

Adamov, S., J. Oskarsson, L. Denby, T. Landelius, K. Hintz,
S. Christiansen, I. Schicker, C. Osuna, F. Lindsten, O. Fuhrer,
and S. Schemm, 2025: Building machine learning limited area models:
kilometer-scale weather forecasting in realistic settings. \emph{ArXiv},
\href{https://arxiv.org/abs/2504.09340}{\underline{https://arxiv.org/abs/2504.09340}}.

Alet, F., I. Price, A. El-Kadi, D. Masters, S. Markou, T. R. Andersson,
J. Stott, R. Lam, M. Willson, A. Sanchez-Gonzalez, and P. Battaglia,
2025: Skillful joint probabilistic weather forecasting from marginals.
\emph{ArXiv},
\href{https://arxiv.org/abs/2506.10772}{\underline{https://arxiv.org/abs/2506.10772}}.

Andrychowicz, M., Espeholt, L., Li, D., Merchant, S., Merose, A., Zyda,
F., Agrawal, S., and Kalchbrenner, N., 2023: Deep learning for day
forecasts from sparse observations. \emph{ArXiv},
\href{https://arxiv.org/abs/2306.06079}{\underline{https://arxiv.org/abs/2306.06079}}.

Ba\~{n}o-Medina, J., A. Sengupta, D. Steinhoff, P. Mulrooney, T. Nipen,
M. Santa-Cruz, Y. Nie, and L. Delle Monache, 2025: A regional
high-resolution AI weather model for the prediction of atmospheric rivers
and extreme precipitation. \emph{npj Clim. Atmos. Sci.}, \textbf{8},
385, \href{https://doi.org/10.1038/s41612-025-01265-9}{\underline{https://doi.org/10.1038/s41612-025-01265-9}}.

Bauer, P., A. Thorpe, and G. Brunet, 2015: The quiet revolution of
numerical weather prediction. \emph{Nature,} \textbf{525}, 47--55.
\href{https://doi.org/10.1038/nature14956}{\underline{https://doi.org/10.1038/nature14956}}.

Bodnar, C., W. P. Bruinsma, A. Lucic, M. Stanley, J. Brandstetter, P.
Garvan, M. Riechert, J. Weyn, H. Dong, A. Vaughan, J. K. Gupta,
K.\ Tambiratnam, A. Archibald, E. Heider, M. Welling, R. E. Turner, P.
Perdikari, 2024: Aurora: a foundation model of the atmosphere.
\emph{ArXiv},
\href{https://arxiv.org/abs/2405.13063}{\underline{https://arxiv.org/abs/2405.13063}}.

Bonev, B., T. Kurth, A. Mahesh, M. Bisson, J. Kossaifi, K. Kashinath,
A. Anandkumar, W. D. Collins, M. S. Pritchard, and A. Keller, 2025:
FourCastNet 3: A geometric approach to probabilistic machine-learning
weather forecasting at scale. \emph{ArXiv},
\href{https://arxiv.org/abs/2507.12144}{\underline{https://arxiv.org/abs/2507.12144}}.

Bostrom, A., and others, 2024: Trust and trustworthy artificial
intelligence: A research agenda for AI in the environmental sciences.
\emph{Risk Analysis}, 44, 1498--1513. DOI: 10.1111/risa.14245.

Chen, L., X. Zhong, F. Zhang, \emph{et al.}, 2023: FuXi: a cascade
machine learning forecasting system for 15-day global weather forecast.
\emph{npj Clim. Atmos. Sci.}, \textbf{6}, 190.
\href{https://doi.org/10.1038/s41612-023-00512-1}{\underline{https://doi.org/10.1038/s41612-023-00512-1}}.

Clough, S.A., M.W. Shephard, E.J. Mlawer, J.S. Delamere, M.J. Iacono,
K.\ Cady-Pereira, S. Boukabara, P.D. Brown, 2005: Atmospheric radiative
transfer modeling: a summary of the AER codes, \emph{J. Quant.
Spectrosc. Radiat. Transfer}., \textbf{91}, 233--244.

Dueben, P. D., and P. Bauer, 2018: Challenges and design choices for
global weather and climate models based on machine learning.
\emph{Geoscientific Model Development}, 11(10), 3999--4009. DOI:
\href{https://doi.org/10.5194/gmd-11-3999-2018}{\underline{https://doi.org/10.5194/gmd-11-3999-2018}}.

Ek, M. B., K. E. Mitchell, Y. Lin, E. Rogers, P. Grunmann, V. Koren, G.
Gayno, and J. D. Tarpley, 2003: Implementation of Noah land surface
model advances in the National Centers for Environmental Prediction
operational mesoscale Eta model, \emph{J. Geophys. Res.}, \textbf{108},
8851, doi:\href{https://doi.org/10.1029/2002JD003296}{\underline{10.1029/2002JD003296}}, D22.

Fowler, L.D., M.C. Barth, and K. Alapaty, 2020: Impact of scale-aware
deep convection on the cloud liquid and ice water paths and
precipitation using the Model for Prediction Across Scales (MPASv-5.2).
\emph{Geosci. Model Dev}., \textbf{13}, 2851--2877,
\href{https://doi.org/10.5194/gmd-13-2851-2020}{\underline{https://doi.org/10.5194/gmd-13-2851-2020}}.

Galarneau, T. J., L. F. Bosart, and R. S. Schumacher, 2010: Predecessor
rain events ahead of tropical cyclones. \emph{Mon. Wea. Rev.},
\textbf{138}, 3272--3297,
\href{https://doi.org/10.1175/2010MWR3243.1}{\underline{https://doi.org/10.1175/2010MWR3243.1}}.

Gneiting, T., and A. E. Raftery, 2007: Strictly proper scoring rules, prediction, and estimation. \emph{J. Amer. Stat. Assoc.}, \textbf{102}, 359--378,
\href{https://doi.org/10.1198/016214506000001437}{\underline{https://doi.org/10.1198/016214506000001437}}.

Guan, H., and Coauthors, 2022: GEFSv12 reforecast dataset for supporting
subseasonal and hydrometeorological applications. \emph{Mon. Wea. Rev.},
\textbf{150}, 647--665,
\href{https://doi.org/10.1175/MWR-D-21-0245.1}{\underline{https://doi.org/10.1175/MWR-D-21-0245.1}}.

Hamill, T. M., and others, 2021: The reanalysis for the Global Ensemble
Forecast System, version 12. \emph{Mon. Wea. Rev}., \textbf{150}, 59--79.

Hamill, T. M., 2026: High-resolution calibrated probabilistic hourly
precipitation from a deterministic forecast. \emph{ArXiv},
\href{https://arxiv.org/abs/2608.12685}{\underline{https://arxiv.org/abs/2608.12685}}.

Han, T., S. Guo, F. Ling, K. Chen, J. Gong, J. Luo, J. Gu, K. Dai, W.
Ouyang, L. Bai, 2024: FengWu-GHR: Learning the kilometer-scale
medium-range global weather forecasting. \emph{ArXiv},
\href{https://arxiv.org/abs/2402.00059}{\underline{https://arxiv.org/abs/2402.00059}}.

Heinzeller, D., M.G. Duda, and H. Kunstmann, 2016: Towards
convection-resolving, global atmospheric simulations with the Model for
Prediction Across Scales (MPAS) v3.1: an extreme scaling experiment.
\emph{Geosci. Model Dev}., \textbf{9}, 77--110,
doi:10.5194/gmd-9-77-2016.

Hersbach H., Bell B, Berrisford P, et al., 2020: The ERA5 global
reanalysis. \emph{Quart J Royal Meteor. Soc}., \textbf{146}, 1999--2049.
\url{https://doi.org/10.1002/qj.3803}.

Hong, S.-Y. and J.-O. J. Lim, 2006: The WRF single-moment 6-class
microphysics scheme (WSM6). \emph{J. Korean Met. Soc.}, \textbf{42}(2),
129--151.

Hu, Z., A. Subramaniam, N. Keen, T. Ge, J. Pathak, M. S. Abbas,
S. Ravuri, K. Kashinath, N. Mahfouz, P. Caldwell, M. Pritchard,
and N. Brenowitz, 2026: Scaling storm-resolving atmospheric AI
simulation to the entire planet. \emph{ArXiv},
\href{https://arxiv.org/abs/2606.31248}{\underline{https://arxiv.org/abs/2606.31248}}.

Ingstad, M. S., M. C. A. Clare, O. Ersland, V. Gahlen,
H. H. Haugen, O. Miralles, E. M. Nordhagen, T. N. Nipen,
I. A. Seierstad, J. B. Bremnes, M. Maier-Gerber,
Z. Ben Bouall\`{e}gue, H. Cook, C. Lessig, G. Mertes,
C. O'Brien, F. Pinault, A. Prieto Nemesio, and M. Chantry,
2026: HourGlass: A probabilistic data-driven temporal downscaler
for global and regional weather forecasting. \emph{ArXiv},
\href{https://arxiv.org/abs/2607.11457}{\underline{https://arxiv.org/abs/2607.11457}}.

Klemp, J. B., 2011: A terrain-following coordinate with smoothed coordinate
surfaces. \emph{Mon. Wea. Rev.}, \textbf{139}, 2163--2169,
\href{https://doi.org/10.1175/MWR-D-10-05046.1}{\underline{https://doi.org/10.1175/MWR-D-10-05046.1}}.

Klöwer, M., Razinger, M., Dominguez, J.J., Dueben, P. D. and
Palmer, T. M., 2021: Compressing atmospheric data into its real
information content. \emph{Nat Comput Sci} \textbf{1}, 713--724.
doi:\href{https://doi.org/10.1038/s43588-021-00156-2}{\underline{https://doi.org/10.1038/s43588-021-00156-2}}.

Kochkov, D., J. Yuval, I. Langmore, P. Norgaard, J. Smith, G. Mooers, M.
Klower, J. Lottes, S. Rasp, P. Duben, S. Hatfield, P. Battaglia, A.
Sanchez-Gonzalez, M. Willson, M. P. Brenner, and S. Hoyer, 2024: Neural
general circulation models for weather and climate. \emph{ArXiv},
\url{https://arxiv.org/abs/2311.07222}.

Kravtsov, S., P. Roebber, T. M. Hamill, and J. Brown, 2022: Objective methods for thinning the frequency of reforecasts while meeting post-processing and model validation needs. \emph{Wea. Forecasting}, \textbf{37}, 727--748,
\href{https://doi.org/10.1175/WAF-D-21-0162.1}{\underline{https://doi.org/10.1175/WAF-D-21-0162.1}}.

Lam, R., A. Sanchez-Gonzalez, and others, 2023: GraphCast: Learning
skillful medium-range global weather forecasting. \emph{ArXiv},
\href{https://arxiv.org/abs/2212.12794}{\underline{https://arxiv.org/abs/2212.12794}}.

Lancelin, A., A. Wikner, L. Dubus, C. Le Priol, D. S. Abbot, F. Bouchet, P. Hassanzadeh, and J. Weare, 2026: AI-boosted rare event sampling to characterize extreme weather. \emph{Phys. Rev. Lett.}, \textbf{137}, 064201,
\href{https://doi.org/10.1103/b1gc-9c2q}{\underline{https://doi.org/10.1103/b1gc-9c2q}}.

Lang, S., M. Alexe, M. Chantry, J. Dramsch, F. Pinault, B. Raoult, Z.
Ben Bouall\`{e}gue, M. Clare, C. Lessig, L. Magnusson, A.P. Nemesio, 2023:
AIFS: a new ECMWF forecasting system. \emph{ECMWF Newsletter,}
\textbf{178}, Winter 2023--2024. doi: 10.21957/1a8466ec2f.

Larsson, E., J. Oskarsson, T. Landelius, and F. Lindsten, 2025:
Diffusion-LAM: Probabilistic limited area weather forecasting with
diffusion. \emph{ArXiv},
\href{https://arxiv.org/abs/2502.07532}{\underline{https://arxiv.org/abs/2502.07532}}.

Li, L., R. Carver, I. Lopez-Gomez, F. Sha, J. Anderson, 2023: SEEDS:
emulation of weather forecast ensembles with diffusion models.
\emph{ArXiv},
\href{https://arxiv.org/abs/2306.14066}{\underline{https://arxiv.org/abs/2306.14066}}.

Mardani, M., N. Brenowitz, Y. Cohen, J. Pathak, C.-Y. Chen, C.-C. Liu,
A. Vahdat, K. Kashinath, J. Kautz, M. Pritchard, 2023: Generative
residual diffusion modeling for km-scale atmospheric downscaling.
\emph{ArXiv},
\href{https://arxiv.org/abs/2309.15214v2}{\underline{https://arxiv.org/abs/2309.15214v2}}.

Mlawer, E.J., S.J. Taubman, P.D. Brown, M.J. Iacono and S.A. Clough,
1997: RRTM, a validated correlated-k model for the longwave. \emph{J.
Geophys. Res}., \textbf{102}, 16663--16682.

Nguyen, T., J. Brandstetter, A. Kapoor, J. K. Gupta, A. Grover, 2023:
ClimaX: A foundation model for weather and climate. \emph{ArXiv},
\url{https://arxiv.org/abs/2301.10343}.

NOAA Environmental Information Systems Working Group, 2024:
\emph{Statement on NOAA Investment in Deep Learning Numerical Weather
Prediction}. NOAA Science Advisory Board, 6~pp. Available at
\href{https://sab.noaa.gov/wp-content/uploads/EISWG-Statement-on-Deep-Learning-NWP_Final_06-18-2024.pdf}{\underline{https://sab.noaa.gov/\ldots}}.

Park, S.-H., J. B. Klemp and W. C. Skamarock, 2014: A comparison of mesh
refinement in the global MPAS-A and WRF models using an idealized
normal-mode baroclinic wave simulation. \emph{Mon. Wea. Rev}.,
\textbf{142}, 3614--3634. doi:10.1175/MWR-D-14-00004.1.

Pathak, J., S. Subramanian, P. Harrington, S. Raja, A. Chattopadhyay, M.
Mardani, T. Kurth, D. Hall, Z. Li, K. Azizzadenesheli, P. Hassanzadeh,
K. Kashinath, and A. Anandkumar, 2022: FourCastNet: A global data-driven
high-resolution weather model using adaptive fourier neural operators.
\emph{ArXiv},
\href{https://arxiv.org/abs/2202.11214}{\underline{https://arxiv.org/abs/2202.11214}}.

Pathak, J., Y. Cohen, P. Garg, P. Harrington, N. Brenowitz, D. Durran,
M. Mardani, A. Vahdat, S. Xu, K. Kashinath, and M. Pritchard, 2024:
Kilometer-scale convection allowing model emulation using generative
diffusion modeling (StormCast). \emph{ArXiv},
\href{https://arxiv.org/abs/2408.10958}{\underline{https://arxiv.org/abs/2408.10958}}.

Pathak, J., M. S. Abbas, P. Harrington, Z. Hu, N. Brenowitz, S. Ravuri,
A. Carpentieri, J. Leinonen, C. Adams, O. Hennigh, N. Geneva, D. Durran,
and M. Pritchard, 2026: Learning accurate storm-scale evolution from
observations. \emph{ArXiv},
\href{https://arxiv.org/abs/2601.17268}{\underline{https://arxiv.org/abs/2601.17268}}.

Perkins, W. A., A. Kwa, J. McGibbon, T. Arcomano, S. K. Clark,
O. Watt-Meyer, C. S. Bretherton, and L. M. Harris, 2025: HiRO-ACE:
Fast and skillful AI emulation and downscaling trained on a 3-km global
storm-resolving model. \emph{ArXiv},
\href{https://arxiv.org/abs/2512.18224}{\underline{https://arxiv.org/abs/2512.18224}}.

Price, I., A. Sanchez-Gonzalez, F. Alet, T. R. Andersson, A. El-Kadi, D.
Masters, T. Ewalds, J. Stott, S. Mohamed, P. Battaglia, R. Lam, M.
Willson, 2024: GenCast: Diffusion-based ensemble forecasting for
medium-range weather. \emph{ArXiv},
\href{https://arxiv.org/abs/2312.15796}{\underline{https://arxiv.org/abs/2312.15796}}.

Rasmussen, R. M., and Coauthors, 2023: CONUS404: The NCAR--USGS 4-km
long-term regional hydroclimate reanalysis over the CONUS. \emph{Bull.
Amer. Meteor. Soc.}, \textbf{104}, E1382--E1408,
\href{https://doi.org/10.1175/BAMS-D-21-0326.1}{\underline{https://doi.org/10.1175/BAMS-D-21-0326.1}}.

Sandbach, S., J. Thuburn, D. Vassilev, and M. G. Duda, 2015: A
semi-implicit version of the MPAS-atmosphere dynamical core. \emph{Mon.
Wea. Rev}., \textbf{143}, 3838--3855. doi:10.1175/MWR-D-15-0059.1.

Seaber, P.R., Kapinos, F.P., and Knapp, G.L., 1987:
\href{https://pubs.usgs.gov/wsp/2294/plate-1.pdf}{\underline{Hydrologic
Unit Maps}}. U.S. Geological Survey
\href{https://pubs.usgs.gov/wsp/wsp2294/}{\underline{Water-Supply Paper
2294}}, 63~p.

Sha, Y., T. Hertneky, E. Gutmann, S. McGinnis, R. McCrary, L. Xue,
D. J. Gagne II, K. Newman, and A. Newman, 2026: AI-based regional
emulation for kilometer-scale dynamical downscaling. \emph{ArXiv},
\href{https://arxiv.org/abs/2602.18646}{\underline{https://arxiv.org/abs/2602.18646}}.

Skamarock, W. C., J. B. Klemp, M. G. Duda, L. D. Fowler, S. Park, and T.
D. Ringler, 2012: A Multiscale Nonhydrostatic Atmospheric Model Using
Centroidal Voronoi Tesselations and C-Grid Staggering. \emph{Mon. Wea.
Rev.}, \textbf{140}, 3090--3105,
\href{https://doi.org/10.1175/MWR-D-11-00215.1}{\underline{https://doi.org/10.1175/MWR-D-11-00215.1}}.

Vannitsem, S., D. S. Wilks and J. W. Messner, 2019: \emph{Statistical
Postprocessing of Ensemble Forecasts}. Elsevier Press, DOI:
\href{https://doi.org/10.1016/C2016-0-03244-8}{\underline{https://doi.org/10.1016/C2016-0-03244-8}},
347~pp.

Wang, W., 2022: Forecasting convection with a ``scale-aware'' Tiedtke
cumulus parameterization scheme at kilometer scales. \emph{Wea.
Forecasting}, \textbf{37}, 1491--1507,
\href{https://doi.org/10.1175/WAF-D-21-0179.1}{\underline{https://doi.org/10.1175/WAF-D-21-0179.1}}.

Weyn, J.A., D. R. Durran, and R. Caruana, 2020: Improving data-driven
global weather prediction using deep convolutional neural networks on a
cubed sphere. \emph{J. Adv. in Modeling Earth Sys}., \textbf{12}, DOI:
\href{https://doi.org/10.1029/2020MS002109}{\underline{https://doi.org/10.1029/2020MS002109}}.

Weyn, J.A., D. R. Durran, R. Caruana, and N. Cresswell-Clay, 2021:
Sub-seasonal forecasting with a large ensemble of deep-learning weather
prediction models. \emph{J. Adv. in Modeling Earth Sys.}, \textbf{13},
e2021MS002502. DOI: 10.1029/2021MS002502.

Zhu, Y., and R. E. Newell, 1994: Atmospheric rivers and bombs.
\emph{Geophys. Res. Letters}, \textbf{21}(18), 1999--2002.
doi:\href{https://doi.org/10.1029/94GL01710}{\underline{10.1029/94GL01710}}.

\end{hangparas}

\end{document}